%% file: purification-ton.tex
\documentclass[journal]{IEEEtran}

\usepackage{times}

\usepackage{graphicx}
\usepackage{color}
\usepackage{dcolumn}
\usepackage{amsmath}
\usepackage{bm}
\usepackage{graphics}
\usepackage{subfigure}
\usepackage{latexsym}
\usepackage{color}
\usepackage{cite}
\usepackage{hyperref}

\def\comment#1{}

\begin{document}


\title{System Design for a Long-Line Quantum Repeater}

\author{Rodney~Van~Meter,~\IEEEmembership{Member,~IEEE,} Thaddeus
  D. Ladd, W.J.~Munro, and Kae~Nemoto\thanks{R. Van Meter is with Keio
    University and National Institute of Informatics, Tokyo, Japan
    (NII) (email: rdv@sfc.wide.ad.jp).  K. Nemoto is with NII (email:
    nemoto@nii.ac.jp).  T. D. Ladd is with NII and with Stanford
    University (email: tdladd@gmail.com).  W.J. Munro is with
    Hewlett-Packard Laboratories, Bristol, UK and with NII (email:
    bill.munro@hp.com).}}



\maketitle

\begin{abstract}
We present a new control algorithm and system design for a network of
quantum repeaters, and outline the end-to-end protocol architecture.
Such a network will create long-distance quantum states, supporting
quantum key distribution as well as distributed quantum computation.
Quantum repeaters improve the reduction of quantum-communication
throughput with distance from exponential to polynomial.  Because a
quantum state cannot be copied, a quantum repeater is not a signal
amplifier.  Rather, it executes algorithms for quantum teleportation
in conjunction with a specialized type of quantum error correction
called {\em purification} to raise the {\em fidelity} of the quantum
states.  We introduce our {\em banded} purification scheme, which is
especially effective when the fidelity of coupled qubits is low,
improving the prospects for experimental realization of such systems.
The resulting throughput is calculated via detailed simulations of a
long line composed of shorter hops.  Our algorithmic improvements
increase throughput by a factor of up to fifty compared to earlier
approaches, for a broad range of physical characteristics.
\end{abstract}


\IEEEpeerreviewmaketitle




\section{Introduction}
\label{sec:intro}

\PARstart{Q}{uantum} computers exist, and have been used to solve
small
problems~\cite{vandersypen:shor-experiment,gulde03:_implem_deuts_jozsa}.
The range of potential uses includes some important problems such as
Shor's algorithm for factoring large numbers and physical simulations
of quantum systems; for a few applications, quantum computers may
exhibit exponential speedup over classical
computers~\cite{nielsen-chuang:qci,shor:factor,abrams99:_expo_quant_algo}.
However, the engineering challenges of creating large-scale quantum
computers are daunting~\cite{van-meter:qarch-impli,kok-2007-79}, and
current capacities are only up to about 8-12 quantum bits, or {\em
  qubits}~\cite{negrevergne06:_12qubit-bench,haeffner05:qubyte}.
Therefore, some researchers have suggested that networks of small
quantum computers be used to overcome the limitations of individual
machines, creating distributed quantum
systems~\cite{grover97:_quant_telec,cleve1997sqe,cirac97:_distr_quant_comput_noisy_chann,van-meter07:_distr_arith_jetc,yepez01:_type_ii,lloyd:quantum-internet}.
The goals of a quantum network are the same as any classical
distributed system: to connect computational resources, data, or
people so that the resulting system is more valuable than the sum of
its parts.  The distant systems may have access to different data, may
provide different computational capabilities, or may simply increase
total capacity.

The first real-world deployments of quantum networks have already
begun.  The first and most developed application is \emph{quantum key
  distribution} (QKD), which uses a quantum channel and an
authenticated (but not necessarily secret) classical channel to create
shared, secret, random classical bits that can be used as a
cryptographic key~\cite{bennett:bb84}\footnote{Note that QKD does not
  completely solve the security problems created by Shor's quantum
  algorithm for factoring large numbers and finding discrete
  logarithms; Shor impacts public-key encryption (which is used in
  authentication mechanisms) and the Diffie-Hellman key agreement
  protocol.  QKD provides key exchange, but requires
  authentication~\cite{paterson04:why-qkd}.}.  An experimental
metropolitan-area QKD network has been developed and deployed in the
Boston area~\cite{elliott:qkd-net}, and similar efforts are underway
in Japan~\cite{nambu06:_one_quant_key_distr_system} and
Europe~\cite{alleaume07:_secoqc_white}.  Efforts to extend these
networks to wider areas are constrained by loss in the communication
channel, which results in exponential decay in throughput as distance
increases.

When the end points of a connection are far apart, the use of
specialized devices called {\em quantum repeaters} may be
required~\cite{briegel98:_quant_repeater,childress05:_ft-quant-repeater,hartmann06,van-loock06:_hybrid_quant_repeater,ladd06:_hybrid_cqed,yamamoto2003eee}.
A quantum repeater is qualitatively different from a classical signal
amplifier; it does not copy a quantum state or regenerate signal
levels (as this is provably impossible in
general~\cite{wootters:no-cloning}).  Instead, quantum repeaters
transfer quantum data via a distributed quantum algorithm called
\emph{teleportation}~\cite{bennett:teleportation,bouwmeester:exp-teleport,furusawa98},
which allows the transfer of a quantum state via classical
communication.  Experimental progress toward the realization of such
repeaters has recently been reported~\cite{Chin-WenChou06012007,zhao2003ere}.

Teleportation consumes a special form of \emph{entangled state} known
as a \emph{Bell pair}.  In an entangled state, two quantum systems
that may be physically separated share a non-local correlation that
Einstein famously referred to as ``spooky action at a distance''.  QKD
does not directly require entangled states.  However, the distributed
Bell pairs created by repeaters will enable long-distance QKD, and most
other applications of distributed quantum computation will use
distributed Bell pairs as
well~\cite{ekert1991qcb,yepez01:_type_ii,lloyd:quantum-internet,van-meter07:_distr_arith_jetc}.

We would like to have perfect Bell pairs to use for our distributed
computations. Unfortunately, perfect systems do not exist, so we must
concern ourselves with the \emph{fidelity} of quantum states, a metric
we will use to describe how near we are to perfect Bell states.  The
fidelity is defined as the probability that a perfect measurement of
two qubits would show them to be in the desired Bell state.  The
fidelity is reduced by channel loss and imperfect control of qubits,
but it may be improved by a form of error correction called
\emph{purification}~\cite{bennett95:_concen,cirac97:_distr_quant_comput_noisy_chann,maneva2000itp,dur2007epa,pan03:_exper-purification}.

The primary contribution of this paper is the introduction of the
\emph{banded purification} algorithm, which improves the utilization
of physical and temporal resources in a network of repeaters.  Our
simulations show that banded purification will improve performance by
up to a factor of fifty compared to prior schemes.  Banding restricts
purification to using Bell pairs of similar fidelity, in order to
improve both the probability of success of the purification and the
resulting boost in fidelity when it succeeds.  We have characterized
expected gains for some system engineering trade-offs.  Our results
increase the performance of the system and relax hardware constraints,
improving the prospects for experimental implementation of wide-area
quantum networks.  We also provide a description of repeater operation
as a network-centric layered protocol model, outlining the messages
transferred, the need for layers to share an addressing scheme for
qubits and the repeaters themselves, and buffer management.

Section~\ref{sec:repeaters} presents the basic operation of quantum
repeaters.  Section~\ref{sec:stack} outlines a layered protocol
architecture to support these operations.  Section~\ref{sec:banded}
describes prior work in scheduling purification, then presents our
banded algorithm.  Our simulation results are detailed in
Section~\ref{sec:sims}, showing the improvement in performance using
banding, as well as the hardware constraints and trade-offs we have
identified.  We conclude in Section~\ref{sec:conclusion}.

\section{Quantum Repeater Basics}
\label{sec:repeaters}

A network of quantum repeaters supports distributed quantum
computation by creating high-fidelity end-to-end Bell pairs.  Once
completed, these pairs can then be used to teleport application data,
which is generally too valuable to risk in the error-prone process of
hop-by-hop teleportation.  Section~\ref{sec:intro} identified the
three functions that a network of quantum repeaters must provide: a
basic entangling mechanism, and the two distributed algorithms,
purification and teleportation, which transform large numbers of
short-distance, low-fidelity Bell pairs into smaller numbers of
long-distance, high-fidelity pairs.

A quantum repeater, which we also call a \emph{station}, is a small,
special-purpose quantum computer, holding a few physical qubits that
it can couple to a transmission medium.  The hardware provides the
basic capability of creating short-distance, low-fidelity (``base
level'') Bell pairs via a physical entanglement
mechanism~\footnote{When we speak of the ``creation'' and
``destruction'' of Bell pairs, we are referring the \emph{state} of
two qubits in separate repeaters; the physical qubits in the repeaters
are not physically created or destroyed.}; the other two functions
require classical communications and computation, and local quantum
operations.

In classical packet-switched networks, an in-flight packet consumes
resources such as buffer space, computation, and bandwidth only at its
current location (modulo end-to-end reliable delivery considerations,
such as TCP).  Quantum repeaters, in contrast, involve widely
distributed quantum computation; each station participates repeatedly
in building an end-to-end distributed Bell pair, through purification
and the use of teleportation known as \emph{entanglement swapping}.

In this section, we first describe how the base-level Bell pairs are
created over a single hop, then how Bell pair fidelity is improved at
all distances by using purification.  With this background, we turn to
teleportation and entanglement swapping.

\subsection{Bell Pair Creation}
\label{sec:epr-creation}

Over distances greater than a few millimeters, the creation of Bell
pairs is mediated by photons, which may be sent through free space or
over a waveguide such as optical fiber.  Schemes for Bell-pair
creation may be divided very crudely into those that use very weak
amounts of light -- single photons, pairs of photons, or laser pulses
of very few
photons~\cite{childress05:_ft-quant-repeater,childress2006ftq,cirac1997qst,duan04:_scalab,van-Enk:PhysRevLett.78.4293,duan2001ldq}
-- and those that use laser pulses of many photons, which are called
\emph{qubus}
schemes~\cite{munro05:_weak,spiller05:_qubus,ladd06:_hybrid_cqed,van-loock06:_hybrid_quant_repeater}.
Qubus repeaters, also known as ``hybrid'' repeaters because they
utilize some analog classical characteristics of light in conjunction
with the digital characteristics of qubits, are currently being
developed by a multi-institution collaboration involving the authors.
The methods of creating photons, performing entangling operations, and
making measurements are different in each type of repeater, but at
the level relevant for this paper the architectures are similar.

At the physical level, the relationship between the probability of
successfully creating a Bell pair and the fidelity of the created pair
is complex.  Only Bell pairs with fidelity bounded well above 0.5
contain useful amounts of entanglement; as the fidelity degrades
toward 0.5, we become unable to make use of the pair.  Using the qubus
scheme, the probability of successfully creating a Bell pair is high,
but even when the operation succeeds the fidelity of the created Bell
pair is low (these two parameters represent an engineering tradeoff we
will not discuss here).  For the parameter settings we have chosen,
corresponding to the qubus scheme, Bell pairs are created with
fidelities of 0.77 or 0.638 for 10km and 20km distances, respectively,
and the creation succeeds on thirty-eight to forty percent of the
attempts~\cite{ladd06:_hybrid_cqed}.  Methods for Bell pair creation
that utilize single photons have much lower success probabilities, but
create very high-fidelity Bell pairs when they do succeed.

\begin{figure}
\centerline{\scalebox{0.5}{\hbox{
\input{pure-round-trips.pstex_t}}}}
\caption{Messaging sequence for the lowest level of Bell pair creation
  and purification.}
\label{fig:pure-round-trips}
\end{figure}
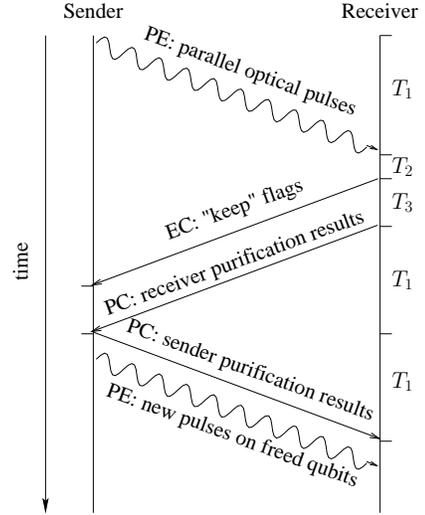

Figure~\ref{fig:pure-round-trips} shows the message sequence for
creating base-level entangled pairs.  The wavy lines in the figure
(labeled PE, for Physical Entanglement) indicate the optical pulses
that interact directly with the qubits, while the straight lines are
classical communication.  At the sender, an optical pulse is entangled
with each separate physical qubit, then multiplexed into the
long-distance fiber.  The pulses are very short compared to the
propagation delay of tens to hundreds of microseconds ($T_1$ in the
figure), so we can treat the pulses as effectively being
instantaneous.  Upon arriving at the receiver, the pulses are
demultiplexed, and an attempt is made to entangle each one with a free
qubit.  Certain properties of the pulse are then
measured~\cite{childress05:_ft-quant-repeater,cirac1997qst,duan04:_scalab,van-Enk:PhysRevLett.78.4293}.
The measurement results tell us if the entangling operation succeeded.
If so, we have created a {\em Bell pair}, entangling a qubit at the
sender with a qubit at the receiver.  The receiver prepares ACK/NAK
``keep'' flags for each qubit and sends them back to the sender,
letting the sender know which operations succeeded.  This measurement
and flag preparation is $T_2$ in the figure and the return message is
labeled EC (Entanglement Control).

\subsection{Purification}
\label{sec:purification}

If the two stations have successfully created more than one Bell pair,
they can next begin the distributed quantum computation known as {\em
purification}.  Purification raises the fidelity of a Bell pair,
essentially performing error correction on a test pattern, taking
advantage of the specially-prepared initial state of the qubits.
Purification takes two Bell pairs and attempts, via local quantum
operations and classical communication, to combine them into one
higher-fidelity pair, an operation that takes time $T_3$ in
Figure~\ref{fig:pure-round-trips}.

Two facets of purification determine its efficiency: the quantum
algorithm used on each pair of Bell pairs, for which there are several
methods~\cite{bennett1996pne,PhysRevLett.77.2818,dehaene2003lpp}, and
the \emph{scheduling}~\cite{dur:PhysRevA.59.169}.  Scheduling chooses
which pairs to purify with each other, and has an enormous impact on
the physical resources required and the rate at which the fidelity of
a Bell pair grows. We will discuss scheduling in detail in
Section~\ref{sec:banded}.  The quantum algorithm used on each pair may
be chosen to be the same regardless of each pair's history, as in
Refs. \cite{bennett1996pne} and \cite{PhysRevLett.77.2818}, but
additional efficiency is gained by tracking the noise accumulated in
each pair as it has developed in the repeater and changing the
algorithm appropriately.  If the noise of the two input Bell pairs is
known, one of a small, finite set of possible algorithms may be chosen
which minimizes the noise of the resulting purified
pair~\cite{dehaene2003lpp}.  We use such an approach for our offline
simulations, assuming the noise expected from qubus-based
hardware~\cite{ladd06:_hybrid_cqed}.  Such an approach is also
possible in real time, but since we cannot directly measure the
quantum parts of the system without disturbing the quantum state, the
quantum state must be tracked by simultaneous classical calculations
identical to our simulations.

Purifying two pairs always destroys one Bell pair and returns its
physical resources to the pool of free qubits.  If the operation
fails, both pairs are freed for reuse, but if the operation succeeds,
the resulting higher-fidelity pair is either kept to await more
purification (if the target fidelity for this distance has not yet
been reached) or is passed to the next higher level in the protocol
stack.  $T_2$ and $T_3$ are both small compared to $T_1$, so we will
ignore them in this paper.

\subsection{Teleportation and Swapping}
\label{sec:teleport-and-swap}

\begin{figure}
\centering
\subfigure[Operations in entanglement swapping.]{
\includegraphics[width=9cm]{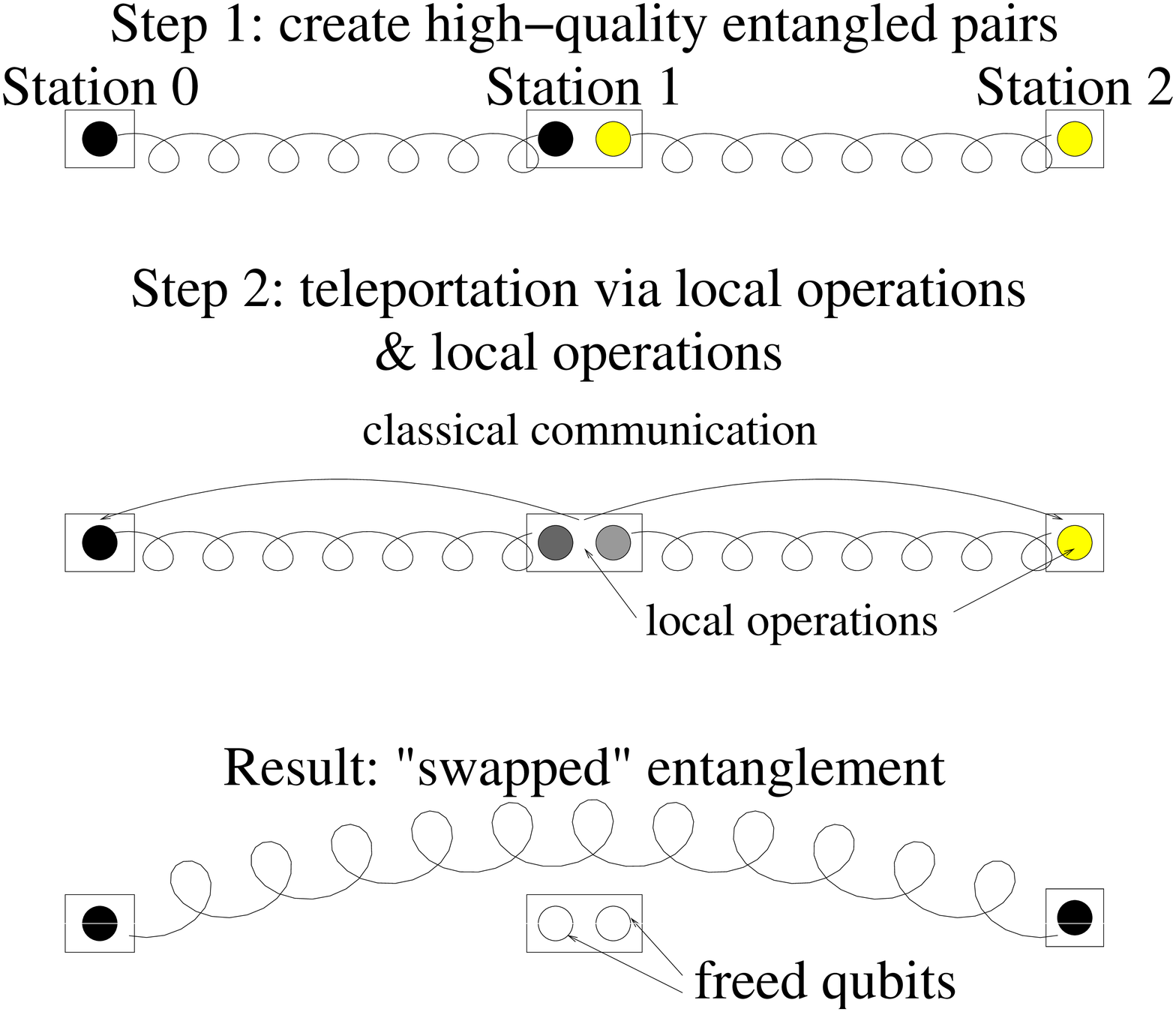}
\label{fig:swapping-ops}}
\subfigure[Three levels of entanglement swapping.]{
  \includegraphics[width=8cm]{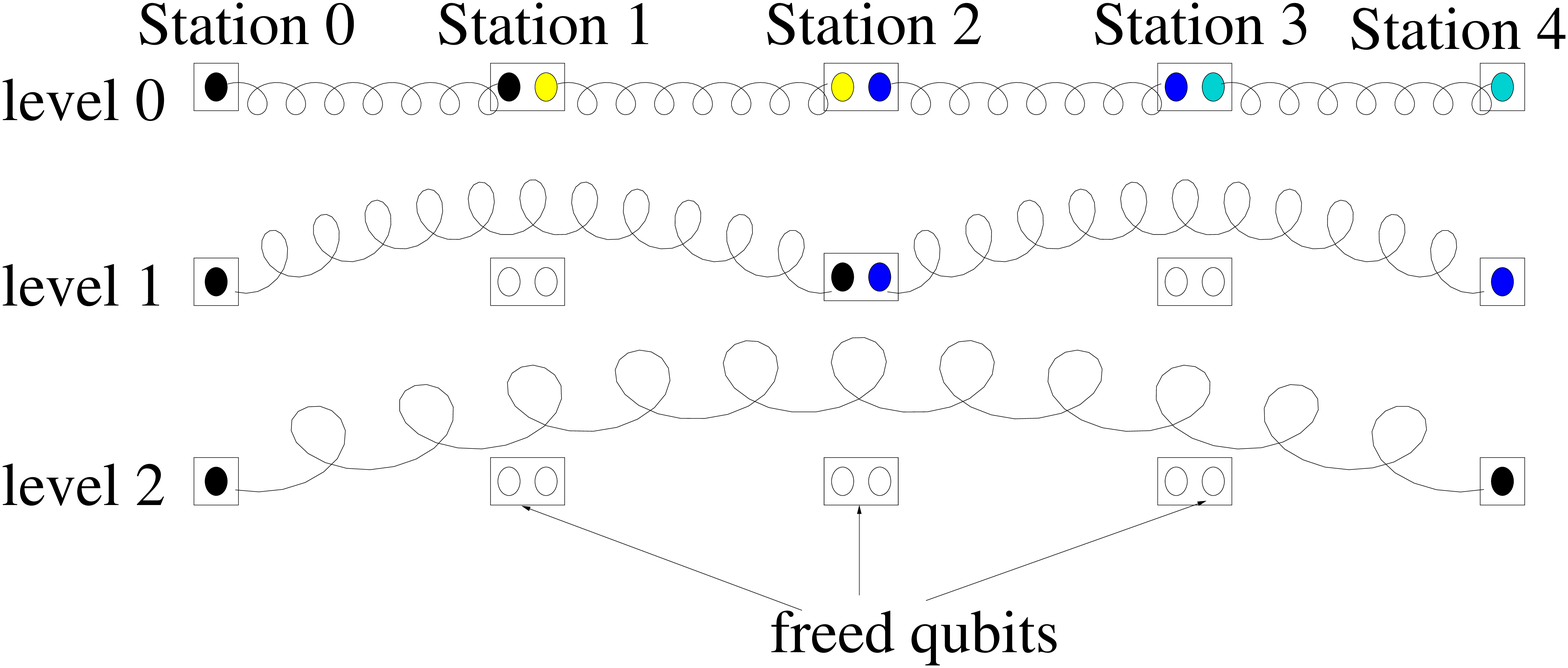}
\label{fig:swapping-3-levels}}
\caption{Entanglement swapping.  Spiral lines represent distributed
  Bell pairs, and straight lines are classical communication.}
\label{fig:swapping}
\end{figure}

The use of teleportation in repeaters, known as entanglement swapping,
lengthens distributed Bell pairs by teleporting the state of one
member of a Bell pair over progressively longer distances, until the
pair stretches from end to end.  Teleportation consumes Bell pairs;
the repeaters are responsible for replenishing their supply of
shorter-distance pairs in order to make the end-to-end Bell pairs.

In teleportation, the state of a qubit is destroyed in one location
and recreated in another.  First, a Bell pair is distributed, with one
member held at the source (Alice) and the other at the destination
(Bob).  The qubit to be teleported (which we will call the data qubit)
is entangled with Alice's member of the Bell pair.  Then both the data
qubit and Alice's Bell qubit are measured.  Each measurement results
in one classical bit, and destroys the quantum state of the qubit.
The two classical measurement results are communicated to Bob, who
then uses them to decide what quantum operations on his Bell qubit
will recreate the original state of the data qubit.

The original creation of the Bell pair must begin with a quantum
operation that entangles two distant qubits, as described in
Section~\ref{sec:epr-creation}, but the teleportation operation itself
requires only local quantum operations and classical communication
between Alice and Bob.

In a system of quantum repeaters, the use of teleportation moves the
state of a single qubit from one station to another.  If the qubit
being teleported is a member of (another) Bell pair, that relationship
is preserved, but one of the end points moves.
Figure~\ref{fig:swapping-ops} illustrates this process, known as {\em
  entanglement swapping}.  A Bell pair spanning nodes 0 and 1
($0\leftrightarrow 1$) and a pair spanning nodes 1 and 2
($1\leftrightarrow 2$) are used to create a single $0\leftrightarrow
2$ pair.  The qubit in Station 1 in step 1 is teleported to Station 2,
lengthening the distance between the end points of the black Bell
pair.  Step 1 is the creation of the base-level Bell pairs.  Step 2 is
local quantum operations at the middle node, including the measurement
of both qubits, followed by classical communication to the end nodes,
then possibly local operations at the destination node to complete the
recreation of the teleported qubit.  This teleportation destroys the
right-hand Bell pair, and frees the two qubits in the middle for
reuse.

In theory, any three nodes with two Bell pairs can use entanglement
swapping, but most designs presented to date assume that a chain of
repeaters will double the distance between end points of the Bell pair
each time swapping is performed, combining two $n$-hop Bell pairs into
one $2n$-hop pair.  Briegel {\em et
  al.}~\cite{briegel98:_quant_repeater} established the use of such a
doubling architecture in early discussions of quantum repeaters, and
showed that performance declines polynomially rather than
exponentially with distance~\footnote{Portions of their analysis apply
  to the splicing of more than two links in each swapping step, but
  they always discuss a regular, exponential growth in the span of
  Bell pairs, and their most detailed analysis uses the doubling
  approach.}.  If purification always succeeds, this logarithmic-depth
combination of links intuitively appears to be optimal, though we know
of no proof of this hypothesis.  Jiang {\em et al.}  have begun
investigating relaxing that constraint dynamically, allowing
neighboring Bell links of any length to combine~\cite{jiang2007oaq}.
This approach is promising for probabilistic systems, and necessary
when physical constraints dictate that the number of hops is not a
power of two.

In the simulations presented here, we assume the use of a basic
doubling architecture for swapping.  We call the number of times
swapping has been performed the ``level'' of the Bell pair, with level
0 being our base Bell pairs at a distance of one hop.  A Bell pair of
level $i$ spans $2^i$ hops.  Figure~\ref{fig:swapping-3-levels} shows
three levels of Bell pairs, representing the state after zero, one,
and two levels of swapping.  In the end, one pair has been stretched
to reach four hops, and the other three Bell pairs present at level 0
have been destroyed and the physical qubits freed for reuse.

\section{Quantum Repeater Protocol Stack}
\label{sec:stack}

\begin{figure*}
\hfill
\subfigure[protocol stack]{
\includegraphics[width=6cm]{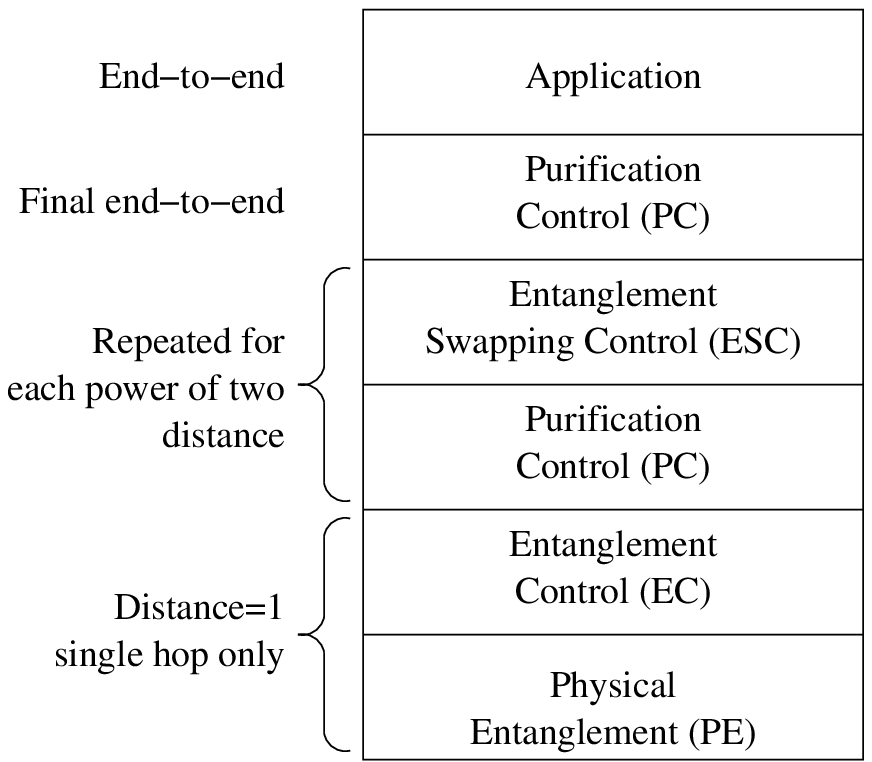}
\label{fig:protocol-stack}
}
\hfill
\subfigure[layer interactions]{
\includegraphics[width=8cm]{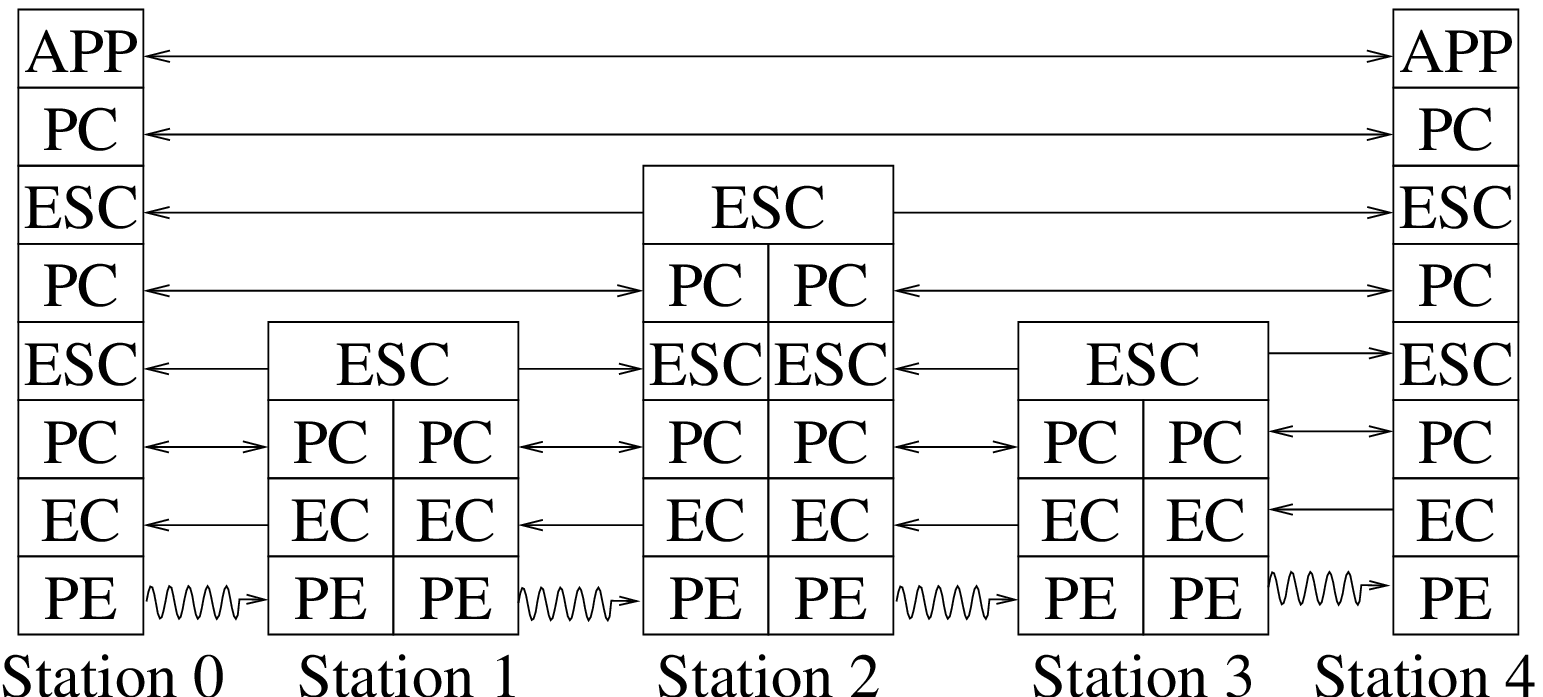}
\label{fig:protocol-interactions}
}
\hfill
\caption{Protocol stack and layer interactions for quantum repeater operation.}
\label{fig:protocol-stacks}
\end{figure*}

\begin{figure}
\center{
\includegraphics[width=8cm]{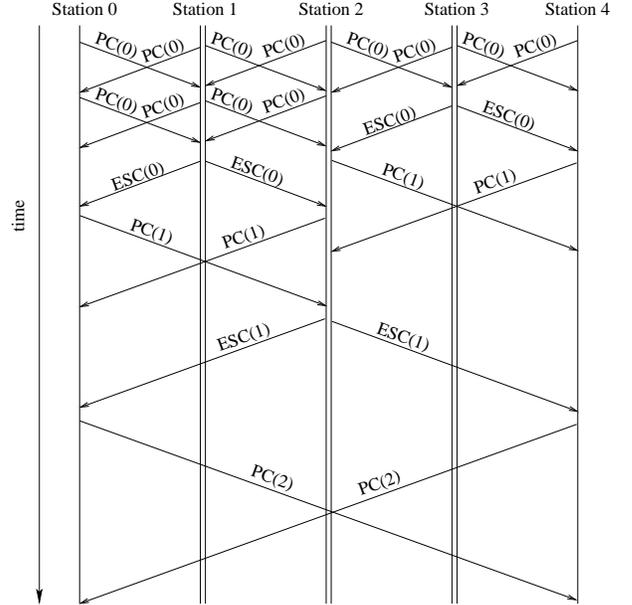}}
\caption{Example message sequence for purification control (PC) and
  entanglement swapping control (ESC).  Numbers in parentheses are the
  level, or distance.}
\label{fig:swapping-msg-seq}
\end{figure}

To give an overview of the processing and message flow in a repeater
network, Section~\ref{sec:repeaters} discussed repeater behavior as an
integrated phenomenon.  However, the actions can be cleanly separated
into a layered protocol stack, as shown in
Figure~\ref{fig:protocol-stack}.  The bottom, {\em physical
entanglement} (PE) layer corresponds to the wavy lines in
Figure~\ref{fig:pure-round-trips}, using strong laser pulses or single
photons to create the shared quantum state between two distant qubits.
The next layer, {\em entanglement control} (EC), consists primarily of
the ``keep'' flags indicating the success or failure of entanglement
attempts.  These bottom two layers operate only across a single hop.
Above these layers reside the {\em purification control} (PC) and {\em
entanglement swapping control} (ESC) layers.  PC consists of a series
of messages indicating the pairs on which purification was attempted,
and the results.  PC must operate at each power of two distance, 1 to
$2^n$, for a $2^n$-hop link.  Figure~\ref{fig:pure-round-trips} shows
the messaging sequence for PE, EC, and the lowest layer of PC.

ESC, which supports the teleportation that splices two Bell pairs to
create one pair that spans a greater distance, must involve three
nodes, as shown in Figure~\ref{fig:swapping} and described in
Section~\ref{sec:teleport-and-swap}.  ESC must inform one of its
partners (generally, the one on the right in the diagram, as we assume
qubits are being teleported left to right) of the results of its local
operations, which are probabilistic.  The right-hand node may need to
perform local operations based on the results received.  The left-hand
node must also be informed of the basic fact of the swapping
operation.  ESC at the middle station unconditionally returns the
qubits just measured to PE for reuse.  The left and right stations
pass control of their qubits to the PC level above the current
ESC, for purification at the new distance.

In normal operation, purification and swapping (PC and ESC) are
repeated at each level until the top, end-to-end level is reached, as
shown in Figures~\ref{fig:protocol-interactions} and
\ref{fig:swapping-msg-seq}.  At that final distance, purification (PC)
may be repeated one more time to create the final end-to-end pair of
the fidelity required by the application.  Of course, purification can
be omitted or repeated at any level, depending on the fidelity of the
Bell pairs.  In Figure~\ref{fig:swapping-msg-seq}, purification at
level 0 is shown happening twice on the left.  The actual timing of
messages may vary somewhat; PC(0) can only be initiated after the
status of qubits has been established by EC, as in
Figure~\ref{fig:pure-round-trips}.  Because the stations run a
deterministic algorithm to select which pairs to purify, PC does not
need to negotiate which operations to perform, only inform its partner
of the outcomes.

When the network is a single line of $N = 2^n$ hops, each station can
easily determine the other stations to which it must build PC and ESC
connections. In a network with a richer topology, this process must
involve routing for the end-to-end connection.  The middle meeting
point of each entanglement swapping level must be identified.  Some
form of source routing or circuit setup will be required, especially
when the number of hops is not a power of two; we defer this problem
to future work.


Both the stations themselves and the qubits they hold must be
addressable.  Because PC and ESC can involve any stations, the control
protocols must be designed to include general station addresses.  EC,
PC, and ESC must also be able to address qubits at both ends of each
connection and to share those addresses with other nodes and protocol
layers.  The addresses can be logical, and a station may relocate its
half of any Bell pair from one internal qubit to another without
notifying its partners, provided it can continue to match incoming and
outgoing messages to the correct qubits.  Once the base Bell pair is
created, the qubits no longer need a direct connection to the long
distance quantum communication channel.

\section{Purification Scheduling}
\label{sec:banded}

Section~\ref{sec:purification} deferred discussion of a critical
point: two stations trying to take a set of lower-fidelity pairs and
create a higher-fidelity pair must decide which pairs to purify.  The
algorithm used determines the physical resources required and the
speed of the convergence to the target fidelity.  Our new banded
purification scheduling algorithm raises the throughput of a given
hardware configuration by a factor of up to fifty, and provides
greater flexibility in hardware configuration.  Before we present
banding, we describe the \emph{symmetric} and \emph{pumping}
scheduling algorithms, then our prior greedy algorithm.

\begin{figure*}
\center{
\includegraphics[width=17cm]{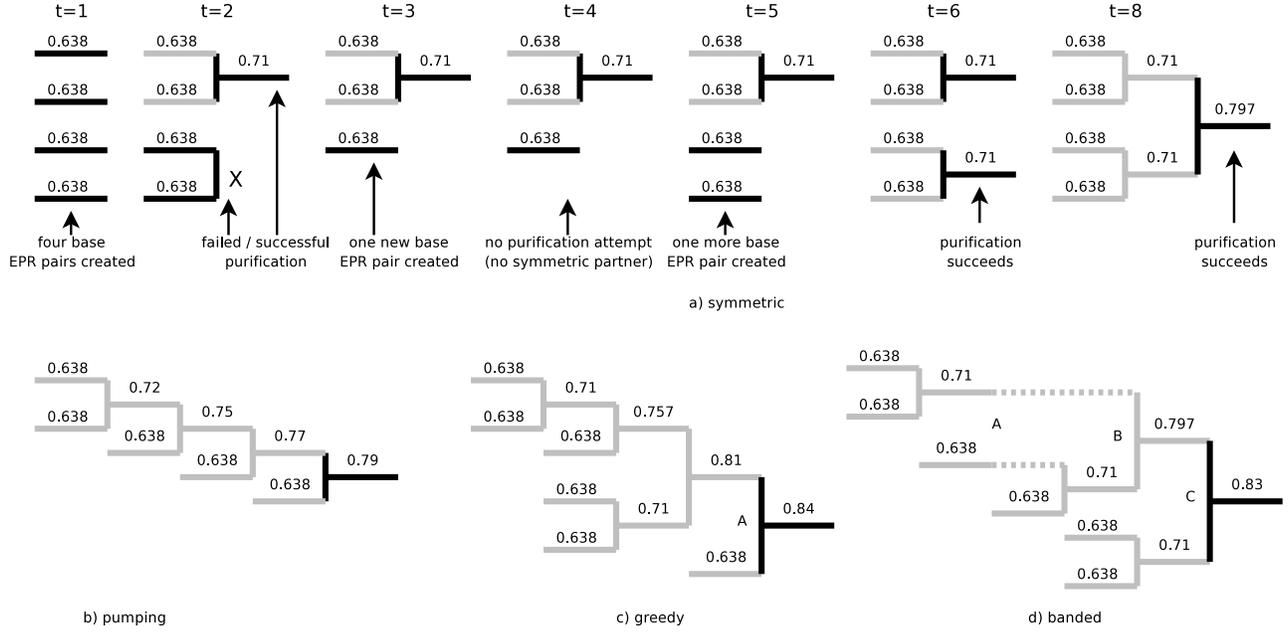}}
\caption{Different purification scheduling algorithms.  Gray bars
  represent the \emph{history} of the pair.  Black horizontal bars
  represent currently entangled Bell pairs.  Numbers show the fidelity
  of the Bell pair.  a) Logical evolution of a symmetrically-grown
  Bell pair.  b) History tree of a Bell pair grown using the
  entanglement pumping algorithm.  c) History tree of a Bell pair
  grown using the greedy algorithm.  d) An example history of the
  evolution of a Bell pair using our new banded purification
  algorithm.  If the boundary between two bands is placed at
  e.g. 0.66, at point A, the pairs 0.71 and 0.638 will not be allowed
  to purify.  Dashed lines represent time that Bell pairs are forced
  to wait for a suitable partner to be created.}
\label{fig:scheduling-trees}
\end{figure*}

Symmetric purification, described by D\"ur \emph{et al.} as schemes A
and B~\cite{dur:PhysRevA.59.169}, requires pairs to attempt to purify
only with other pairs of the same fidelity.
Figure~\ref{fig:scheduling-trees}a shows the evolution and history of
a symmetrically-grown Bell pair.  In the figure, for simplicity,
base-level Bell pairs are created in odd-numbered time steps, and
purification operations are attempted in even-numbered time steps.
(The fidelities in the diagrams in this section are for illustration
only, and are not exact.)  At $t = 4$, the symmetric algorithm would
not attempt to purify the fidelity 0.71 pair with the fidelity 0.638
pair, instead waiting for the development of a second fidelity 0.71
pair at $t = 6$.

\comment{how many nines?}

Symmetric purification would take our starting fidelity of 0.638 to
e.g. a target fidelity of 0.98 after five rounds.  If purification
always succeeded, thirty-two ($2^5$) base-level Bell pairs would be
required: $32\times 0.638 \rightarrow 16\times 0.71 \rightarrow
8\times 0.797 \rightarrow 4\times 0.867 \rightarrow 2\times
0.952 \rightarrow 1\times 0.988$.  Unfortunately, purification is a
state-dependent, probabilistic operation.  When using our starting
state, the first step (0.638 + 0.638) will succeed only 57\% of the
time, while the last step will succeed 92\% of the time.  In total,
symmetric purification actually consumes, on average, more than 450
base-level Bell pairs to make one Bell pair of 0.98 fidelity.

The principal drawbacks to the symmetric algorithm are the inflexible
use of available resources, both time and space (as shown by e.g. the
wait at $t = 4$ in the figure), and the fact that the truly symmetric
history tree is effectively impossible to achieve.  Memory degradation
over time causes two pairs that arrived at different times to have
different fidelities, so forcing exact matches only is impractical.

Entanglement pumping, defined by D\"ur \emph{et al.} as Scheme C and
shown in Figure~\ref{fig:scheduling-trees}b, can be done using only
the minimum two qubits in each
station~\cite{childress05:_ft-quant-repeater}.  The
fidelity of one Bell pair is pumped by purifying with base-level pairs
created using the physical entanglement mechanism.  This scheme uses
physical resources efficiently, but improves the fidelity of
entanglement only slowly; when the fidelity difference between the
base pairs and the final target fidelity is large, pumping is
ineffective.

Previous work~\cite{ladd06:_hybrid_cqed} considered a greedy
scheduling algorithm for purification scheduling: at each time step,
all available resources are purified, never deferring immediate
actions in favor of potential later operations.
Figure~\ref{fig:scheduling-trees}c shows one possible history tree.
When the fidelity of a base Bell pair is high, above $\sim 0.75$ or
so, this scheme works well.  However, when the fidelity is lower,
because of longer distances or loss elsewhere in the system, a greedy
algorithm results in attempts to purify a high-fidelity pair with a
low-fidelity pair, as at the point A in the figure.  Using a
low-fidelity pair both has a lower probability of success and gives
only a small boost in fidelity when it succeeds.  Thus, the effective
floor for the fidelity of base pairs when using the greedy algorithm
is high.

We have seen that the greedy algorithm and entanglement pumping
sometimes match Bell pairs with very different fidelities, resulting
in low probability of success for the purification operation and
giving only a limited boost in fidelity even on success.  The fully
symmetric tree is impractical: it imposes strict minimum hardware
requirements, cannot allocate resources flexibly, and in practice
cannot take into account memory degradation.  A new approach is
required.

We have developed {\em banded} purification to match purification
pairs efficiently but flexibly.  We divide the fidelity space into
several regions, or bands, and only allow Bell pairs within the same
band to purify with each other.  Figure~\ref{fig:scheduling-trees}d
shows a simple example, assuming two bands divided at a fidelity of
0.66.  At the point A, the greedy algorithm would attempt to purify
the 0.638 pair with the 0.71 pair (as shown at A in
Figure~\ref{fig:scheduling-trees}c).  When using banding, the band
boundary at 0.66 prevents those pairs from purifying, and so the
system waits for the creation of another fidelity 0.638 pair, then
purifies the two 0.638 pairs.  If that purification is successful,
resulting in a second 0.71 pair, then purification will be attempted
at point B using the two 0.71 pairs.  At point C, the banding
structure allows the new 0.71 pair to purify with the 0.797 pair,
whereas the symmetric algorithm would block temporarily.  Unlike the
greedy and pumping algorithms, the banded approach treats
high-fidelity pairs as more valuable than low-fidelity pairs, and only
uses them when another similarly high pair is available, making those
operations more likely to succeed and providing a larger boost in
fidelity.

Recall that the purification operations can fail, but their
probability of success increases as the fidelity of the pairs involved
increases.  Any attempt to predict the exact best sequence of
purification operations from a given state, therefore, must take into
account which resources are currently busy, the fidelities of all
available Bell pairs, the probability of success of possible
purification choices, and the probability that currently unentangled
qubits will be successfully entangled in the near future using the
physical entanglement mechanism.

The banded and symmetric algorithms are potentially subject to
deadlock, but the problem is easily solved for the banded algorithm.
If a repeater has e.g. seven qubits and seven bands (or seven rounds
of purification for the symmetric case), one Bell pair could be in
each band.  Each pair would have no possible purification partner, and
no free qubits would be available to create new pairs to add to the
bottom band.  Each swapping level is independent, so the minimum
number of qubits per station must actually be the number of bands
times the number of levels, plus one, for the receive half and send
half of the repeater.  In our simulations, we select a hardware
configuration, then restrict the number of bands used to a number that
will not deadlock.  The symmetric algorithm has no such flexibility.

\section{Simulation Results}
\label{sec:sims}

We have simulated repeater chains for a broad range of the parameters
discussed in prior sections.  The majority of our simulations utilize
banded purification, and the greedy algorithm is simulated for
comparison.  We simulate the quantum mechanics of the physical
interactions and operations, but a large fraction of the code (7,000
lines of C++) and execution time (several weeks on eight 3.0GHz+ Intel
processors) are dedicated to managing the messages that are
transferred station to station.

The metric we use to evaluate quantum networks is the throughput,
measured in Bell pairs per second of a certain fidelity over a given
distance.  We have chosen a target fidelity of 0.98, and simulate for
distances up to 20,000 kilometers.  Unless otherwise specified, the
simulations presented here are for 64 links of 20 kilometers each with
one hundred qubits per station (50 for receive and 50 for send, except
at the end points where all 100 can be used for one direction).  Our
simulations all assume $0.17$ dB/km loss and a signal propagation
speed of $0.7c$, corresponding to telecommunications fiber.  In the
hybrid quantum repeater schemes we simulate, fiber loss translates to
reduced fidelity for a Bell pair, rather than a lower success
rate~\cite{ladd06:_hybrid_cqed,van-loock06:_hybrid_quant_repeater}.
For 20km links at $0.7c$, the one-way latency for signals is just
under 100$\mu$sec, so the ``clock rate'' for these simulations is
about 10kHz.  As noted above, the pulses are very short compared to
the propagation latency, and for the settings we use, entanglement is
successful about forty percent of the time.  With these settings, in
the first time step, each station will attempt to entangle fifty
qubits, successfully creating about twenty base-level Bell pairs on
each link.  In successive time steps, the number of attempts on each
link is capped by the number of available qubits at each station.

Our code is capable of simulating imperfect local gates, but to
isolate the individual factors presented here, the simulations in this
paper assume perfect local quantum operations and memory.  Our
simulations have shown that gate errors of 0.1\% result in about a
factor of two reduction in the performance of the system, with
performance degrading rapidly and a final fidelity of 0.98 being
unattainable with gate errors of 0.3\%.  A complete discussion of
error mechanisms in quantum computing and the current experimental
state of the art is beyond the scope of this paper, but this level of
quality is well beyond what is currently possible; the number 0.1\%
should be viewed as a \emph{target} which experimentalists should
strive to achieve.

As a rough approximation, the gate error rate can be considered to be
the \emph{combination} of both local gate errors and memory errors.
With one-way latency in fiber of approximately 6msec at 1,280km,
memory must be able to retain its state for times on the order of
seconds to meet the above constraint.  Hartmann \emph{et al.} have
recently examined the role of memory errors in quantum
repeaters~\cite{hartmann06}, finding that memory that can successfully
retain a quantum state for about one second can support ultimate
repeater distances of 5-20,000km, albeit it at large cost in resources
and with a cap on the achievable fidelity.  If memory times are
substantially shorter, then local quantum error correction should be
added, which will add substantial additional complexity to the system
design.

For each banded data point in the graphs presented here, extensive
runs over large parameter spaces (up to 800 or so separate sets of
parameter settings) were executed to find a good set of bands, and to
find a good set of thresholds for entanglement swapping at different
distances.  Each data point represents a single run in which 200
end-to-end Bell pairs of final fidelity 0.98 or better are created,
with the exception of a few of the slowest data points, which were
terminated early.  The throughput is calculated by linear regression
to fit a line to the arrival times of the Bell
pairs~\cite{jain:perf-anal}.  Error bars are included for all graphs
except Figure~\ref{fig:variable-hops-latency} but are almost too small
to be seen at many data points; they represent the standard deviation
of the fitted slope for that run.  The coefficient of determination is
above 0.996 for almost every fit except the three data points with the
largest error bars in Figure~\ref{fig:variable-hops}, for which it is
0.95, 0.80, and 0.78.  These fits confirm that despite the stochastic
nature of the quantum operations, the mean arrival rate is constant
after the initial transient startup latency.  Runs of fewer than 200
Bell pairs were found to have unacceptably large variability.  Data,
log files, and parameter settings for all runs are available from the
authors.

First we analyze the performance of the greedy algorithm, then present
our primary results, comparing the throughput of greedy and banded
purification.  We backtrack to explain how bands are selected, then
compare several options for setting the fidelity target at each
swapping level.  The final two subsections explore the hardware
configuration, assessing the importance of the number of qubits per
station and the trade-off of distance versus repeater size.

\subsection{Greedy Algorithm}

\begin{figure}
\center{
\includegraphics[width=8cm]{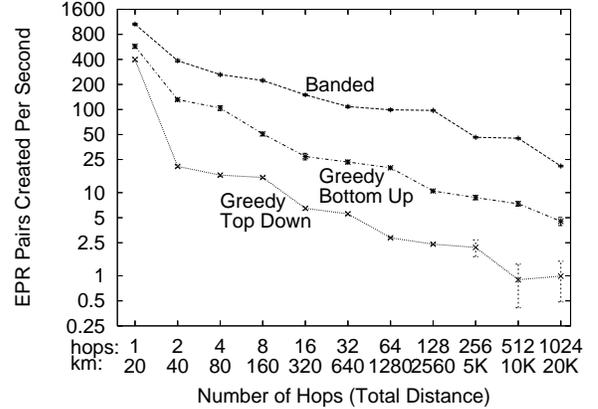}}
\caption{Throughput versus distance for the banded algorithm using
  five bands, compared to the greedy bottom up and greedy top down
  algorithms.  The final fidelity is 0.98.}
\label{fig:variable-hops}
\end{figure}

The performance of the greedy top down algorithm, corresponding to our
prior work, is the bottom line in
Figure~\ref{fig:variable-hops}~\cite{ladd06:_hybrid_cqed}.  Throughput
in end-to-end Bell pairs created per second is plotted against
distance.  The X axis is labeled with both the number of hops and
total distance in kilometers; the rightmost point of 1,024 hops or
20,000 kilometers corresponds roughly to the distance halfway around
the world.

For the greedy top down algorithm, throughput is about 21 Bell
pairs/second for two hops, and declines to almost exactly 1 Bell
pair/second for 1,024 hops.  The decline shows a distinct stair-step
structure, caused by the discrete nature of purification and our
choice to purify until a final fidelity of 0.98 is reached.  At a
particular length, a certain number of purification steps is required
to achieve the final fidelity.  As the number of hops increases, the
same number of purification steps may continue to serve, until the
fidelity drops below the target and an additional round of
purification must be added.  When this happens, the performance drops
by roughly a factor of two, as two high-quality pairs up near the
target are required.

The greedy algorithm sorts the Bell pairs by fidelity, and pairs them
starting with the two highest-fidelity pairs.  We discovered that
pairing beginning from the bottom of the list, which we term greedy
bottom up, increases performance by a factor of three to eight, as the
middle curve in Figure~\ref{fig:variable-hops} shows.  We attribute
this improvement to increased conservatism on the use of the
highest-fidelity pair.  Beginning at the bottom will bring other pairs
up toward the fidelity of the highest pair, perhaps even surpassing
it, but first risking the failure of lower-fidelity pairs which have
cost less to build.

At the left hand edge of the graph, the greedy top down algorithm
declines from 400 pairs/second for one hop to 21 for two hops, almost
a factor of twenty worse.  For this graph, our hardware is assumed to
have one hundred qubits per station.  For one hop, all one hundred
qubits can directly connect to qubits at the far end.  For two hops,
the middle station must split the use of its one hundred qubits, fifty
for the left-hand link and fifty for the right-hand link.  The
difference is due to more efficient purification pairings as the
number of available qubits grows.  This effect is assessed in more
detail in Section~\ref{sec:qubits-per-station}.

\subsection{Banded Performance v. Total Distance}

\begin{figure}
\center{
\includegraphics[width=8cm]{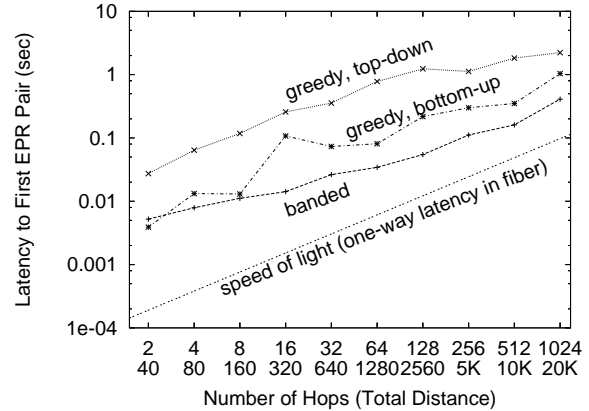}}
\caption{Startup latency versus distance for the banded, greedy bottom
  up, and greedy top down algorithms.}
\label{fig:variable-hops-latency}
\end{figure}

The top line in Figure~\ref{fig:variable-hops} graphs the performance
of our banded algorithm.  Throughput starts at 1060 Bell pairs/second
for one hop, plateaus at about 100 for 32 to 128 hops, then declines to
20 pairs/second for 1,024 hops.  Due to the stair-step behavior, the
benefit compared to the greedy top down algorithm varies from a factor
of fifteen to a factor of fifty, with the advantage growing unevenly
as distance increases.  Compared to the greedy bottom up algorithm,
banded is 2.5 to 9.3 times better, also increasing unevenly with
distance.

Entanglement pumpting and symmetric scheduling are not shown in the
figure.  Entanglement pumping cannot effectively create pairs of
fidelity 0.98 with our starting fidelity of 0.638.  For the particular
configuration shown here, the symmetric algorithm would perform
similarly to banding.  However, as noted in Section~\ref{sec:banded},
the fully symmetric algorithm cannot be achieved in practice.

An important question is whether band structure changes when the total
distance (number of hops) is increased.  If the band structure does
not change, then we can simulate short lines, and apply the simulation
results directly to much longer lines, dramatically reducing the
amount of computation time needed in simulations.  Likewise, in
real-world operational environments, distance-independent system
controls would be a boon.  Unfortunately, our simulations have shown
that the banding structure does vary somewhat at different distances.
The performance for nearby banding structures can be a factor of two
worse, meaning that a careful search is necessary for each specific
link configuration.

Because the Bell pairs created are a generic resource that do not
initially carry application data, the normal operation mode for the
system will be steady-state, continuous operation, buffering prepared
Bell pairs to the extent possible during times when applications are
not consuming them.  As noted in Section~\ref{sec:intro}, the
distributed nature of repeater operations means that there is no true
``in flight'' time for a qubit.  Nevertheless, a quick look at the
latency to start up the system is in order.
Figure~\ref{fig:variable-hops-latency} shows the latency from the time
the system is started until the first end-to-end Bell pair is created.
The values graphed are the latency for the first Bell pair for each of
the runs in Figure~\ref{fig:variable-hops}.  For the banded algorithm,
start-up latency is about fifteen times the one-way latency for two
hops, declining to about four times the latency for 1,024 hops.

\subsection{Finding the Bands}
\label{sec:band-finding}

\begin{figure}
\center{
\includegraphics[width=8cm]{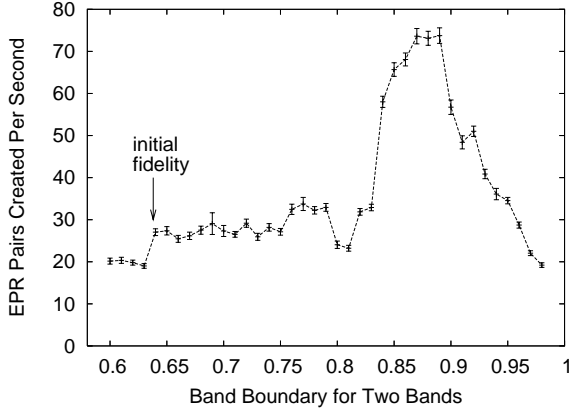}}
\caption{Finding the best band boundary for a 2-band arrangement, for
  64 hops of 20km each.}
\label{fig:bandscan}
\end{figure}

\begin{figure}[t]
\center{
\includegraphics[width=8cm]{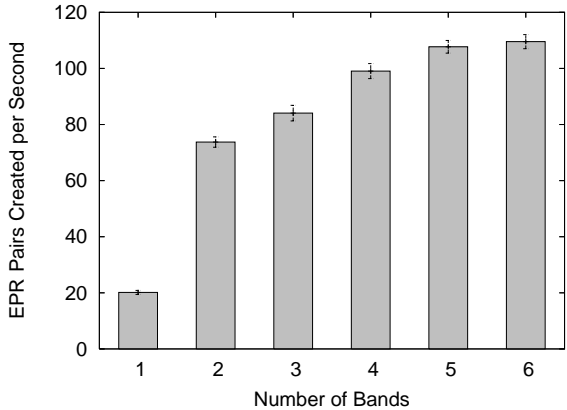}}
\caption{The best band throughput for different numbers of bands, for
  64 hops of 20km each.}
\label{fig:number-of-bands}
\end{figure}

We can theoretically place the boundaries that separate bands at
almost any level.  To determine a placement that gives good
performance, we have performed nearly exhaustive searches over many
possibilities, for configurations with two to six bands.
Figure~\ref{fig:bandscan} shows a two-band setup.  In this figure, we
vary the boundary in steps of 0.01, but in most other graphs the steps
are 0.02 or 0.04.  At the left edge, the division between the two
bands is below the initial threshold of 0.638 generated by our
physical entanglement process, and at the right edge the division is
above the delivery threshold for our final qubits, resulting in the
equivalent of the bottom up greedy algorithm for the first and last
data points.  The performance peaks when the band boundary is
0.87-0.89, showing clearly that the operational imperative is
protecting the high-fidelity pairs from purifying with low-fidelity
pairs.

Increasing the number of bands gives a smooth increase in performance
for up to five bands, which perform nearly 50\% better than two bands.
Figure~\ref{fig:number-of-bands} shows the increase in performance for
increasing numbers of bands.  Moving from one band (equivalent to
greedy bottom up) to two increases performance by more than a factor
of three.  The performance has saturated with six bands; it is not
clearly better than five bands, because the behavior has essentially
been constrained to that of a symmetric tree.  For more than two
bands, the number of simulation runs to cover the space increases
geometrically, so the granularity of our boundary steps is somewhat
larger.  For three bands, for example, we tried all combinations of
boundaries with the lower bound varying 0.60 to 0.95, and the upper
boundary varying from 0.80 to 0.99, in steps of 0.02.

\subsection{Varying Swapping Thresholds}

\begin{figure}
\center{
\includegraphics[width=8cm]{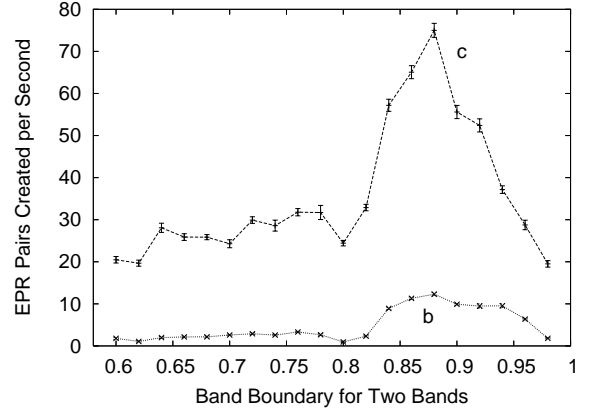}}
\caption{Comparing different distance-swapping thresholds.}
\label{fig:threshold-compare}
\end{figure}

Recall the distinction between the purification bands and thresholds at
different distances: the former governs purification decisions within
PC, while the latter governs the promotion of pairs from PC to ESC for
entanglement swapping at the next-higher distance.  The experiments in
the previous subsections were performed with each of the distance
thresholds set to 0.98.  In this section, we evaluate several possible
sets of thresholds that seem like plausible candidates for good
configurations:

\begin{itemize}
\item[{\bf a.}] 0.9, 0.9, 0.9, 0.9, 0.9, 0.9, 0.98:\\ purify only to
an intermediate fidelity of 0.9 at distance 1, 2, 4, 8, 16, and 32,
then push to the final fidelity of 0.98 at the full distance of 64
hops;
\item[{\bf b.}] 0.98, 0.9, 0.9, 0.9, 0.9, 0.9, 0.98:\\  purify to
  fidelity 0.98 at distance 1, then allow the fidelity to slip as far
  as 0.9 at intermediate distances, before pushing back up to 0.98 at
  64 hops; and
\item[{\bf c.}] 0.98, 0.98, 0.98, 0.98, 0.98, 0.98, 0.98:\\ purify to
  fidelity 0.98 at distance 1, then maintain that fidelity by
  purifying as necessary at each distance.
\end{itemize}

Figure~\ref{fig:threshold-compare} shows clearly that the preferred
method of managing the fidelity of a pair as it hops across the
network is case {\bf c}, purifying to the desired level at distance
one, and maintaining that fidelity at all distances.  Case {\bf a}
proved to perform so poorly that the simulations were unable to
complete.  The other two cases are shown in the figure.

This data supports the intuitive idea that purifying over short
distances will be more efficient than purifying over long distances.
D\"ur \emph{et al.} referred to this approach as maintaining a
``working fidelity''~\cite{dur:PhysRevA.59.169}.  They did not
report on any alternative schemes, but our data confirms that their
approach is correct.

In addition, we investigated several other candidate schemes, all of
which performed worse than maintaining a working fidelity; those
results confirm our findings presented here.

Because the curves for {\bf b} and {\bf c} have the same shape,
despite radically different distance thresholds,
Figure~\ref{fig:threshold-compare} also suggests that changing the
pattern of fidelity thresholds at different distances is {\em
  independent} of the choice of the bands for banded purification.
That is, a good choice of bands should remain good regardless of the
thresholds at various distances.  This fact should allow us to
optimize these two parameters independently for a given physical
configuration.

\subsection{Number of Qubits per Station}
\label{sec:qubits-per-station}

\begin{figure}[t]
\center{
\includegraphics[width=8cm]{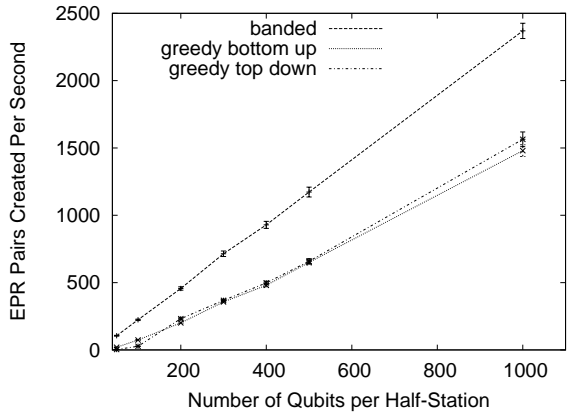}}
\caption{Comparing different numbers of qubits per station, for five
  bands, greedy bottom up, and greedy top down.
Simulations are 64 hops of 20km each.}
\label{fig:number-of-qubits}
\end{figure}

\begin{figure}[t]
\center{
\includegraphics[width=8cm]{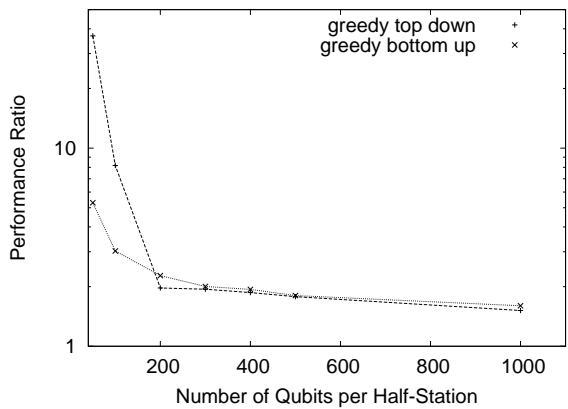}}
\caption{Performance ratio of banded purification compared to the two
  forms of greedy purification, bottom up and top down.}
\label{fig:number-of-qubits-ratio}
\end{figure}

The principal constraint on throughput is the number of qubits per
station.  What will happen as our hardware capabilities grow?  How
should we distribute limited physical resources?  This subsection and
the next address these two important questions.

The throughput achieved for two bands with varying numbers of qubits
per half-station is shown in Figure~\ref{fig:number-of-qubits}.  The
throughput achievable with five-band purification scheduling is linear
in the number of qubits.

Banding is especially valuable in the near term, when station
capacities are expected to be one of the principal engineering
constraints.  The greedy top down algorithm performs poorly with small
numbers of qubits per station.  With larger numbers of qubits in
various stages of development, simply ordering the list and partnering
Bell pairs bottom up will naturally tend to use qubits that are of
similar fidelity, giving similar behavior to banding without a formal
band structure.  Figure~\ref{fig:number-of-qubits-ratio} shows this
effect, with the banded algorithm outperforming the greedy algorithms
at all station sizes, but by a smaller ratio as the station capacity
grows.  With 50 qubits per half-station, the greedy top down algorithm
struggles to meet our fidelity goal of 0.98, and banded performs
thirty-seven times better.

\subsection{Varying Number of Stations}
\label{sec:varying-stations}

\begin{figure}
\center{
\includegraphics[width=8cm]{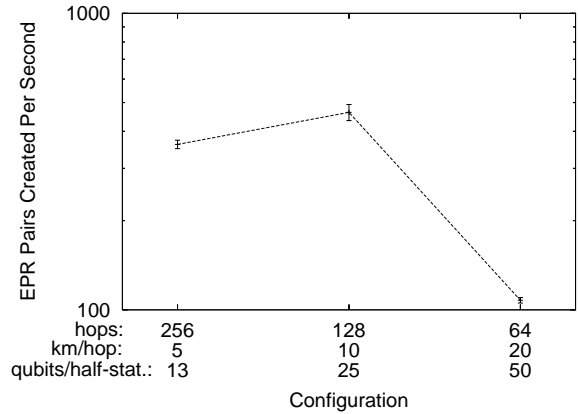}}
\caption{Best throughput for about 6,500 qubits spread over 1,280km.}
\label{fig:fixed-capacity}
\end{figure}

Because physical qubits may be the scarce resource in a repeater
system, it makes sense to ask how best to spread the qubits out along
a link to achieve the maximum throughput.  With the exception of
Section~\ref{sec:qubits-per-station}, most of the experiments
presented so far in this paper have used 64 hops of 20km each with 50
qubits per half-station, but what if we were to split each repeater
and create 128 hops of 10km with 25 qubits per half-station, or 256
hops of 5km with 13 qubits?  These three cases are shown in
Figure~\ref{fig:fixed-capacity}.  As the
number of qubits per half-station decreases, we must restrict the
number of bands in order to avoid deadlock.  For 64 hops and 50
qubits, we can use five bands; for 128 hops and 25 qubits, only three;
and for 256 hops with 13 qubits only a single band.  This fact gives
us an engineering trade-off; bands are especially useful in
lower-fidelity hops, helping to offset the decrease in throughput that
comes from lengthening the hops.

This section has shown some preliminary explorations of this question,
but the space of possibilities is large, and for both long and short
hops, additional factors such as memory errors and local gate errors
will likely play larger roles.  A more complete analysis would require
a combinatorial increase in the number of simulations performed; the
total simulation time of more than half a CPU-year would be multiplied
by the number of swapping thresholds tested for each of seven
thresholds in the configurations above.  We defer more a complete
analysis to future work.

\section{Conclusion}
\label{sec:conclusion}

The banded purification algorithm and hardware parameters presented
here represent a step forward in quantum repeater network design, as
shown by the gains in throughput we report, especially with
intermediate numbers of qubits per station.  Banded purification
provides throughput essentially identical to fully symmetric
purification.  Symmetric purification, however, cannot be achieved in
practical systems, due to memory degradation and the possibility of
deadlock.  These gains are more than hypothetical; the improved
operation at low initial fidelities will assist the first laboratory
experiments of a complete repeater network, which inevitably will
operate at the very edge of a functional system.  Although the basic
concepts of quantum repeaters are simple, physical realizations remain
some years away.  The dynamic behavior is analytically intractable and
the range of engineering parameters broad, making simulation a
valuable tool.  Our simulations have helped us to identify important
hardware constraints and test possible protocols, allowing us to find
improvements that raise performance by a factor of fifteen to fifty
across a broad range of distances and parameters, and to extend the
possible operating range to lower fidelities (down to a fidelity of
below 0.55, compared to greater than 0.7 for prior simulations).

We have laid out a rudimentary architecture for the protocols
necessary to operate a network of repeaters.  We know that buffer
qubits must have addresses at the entanglement control (EC) level and
above.  At the purification control (PC) level and above, stations
must also have addresses.  These addresses must be shareable across
layers of the protocol stack.  Software-selectable characteristics of
the protocols, such as bands and thresholds for promotion to longer
swap distances may be locally-held information only, decided out of
band, or dynamically negotiated through an additional session control
protocol; we defer such design issues until experimental progress
demands.

Banded purification will be useful for quantum system-area networks
(SANs), as well as wide-area quantum networks.  In wide-area quantum
networks, loss is dominated by the length of the fiber.  In SANs for
quantum
multicomputers~\cite{van-meter07:_distr_arith_jetc,van-meter07:_commun_links_distr_quant_comput},
fiber losses will be low, but losses elsewhere in the system (e.g.,
the qubit-fiber coupling or node-to-node switching) will be present,
requiring the use of purification.  Our results will assist the
development of distributed quantum computing systems with node-to-node
distances ranging from a handspan to intercontinental, helping to
usher in the era of quantum computation.


\section*{Acknowledgments}

The authors thank NICT for partial support for this research.  We
thank Kohei M. Itoh, Peter van Loock, John Heidemann and Joe Touch for
valuable discussions.

\bibliography{paper-reviews}

\end{document}

%% file: pure-round-trips.pstex_t
\begin{picture}(0,0)%
\includegraphics{pure-round-trips.pstex}%
\end{picture}%
\setlength{\unitlength}{3947sp}%
\begingroup\makeatletter\ifx\SetFigFont\undefined%
\gdef\SetFigFont#1#2#3#4#5{%
  \reset@font\fontsize{#1}{#2pt}%
  \fontfamily{#3}\fontseries{#4}\fontshape{#5}%
  \selectfont}%
\fi\endgroup%
\begin{picture}(5509,6432)(180,-6373)
\put(4951,-1111){\makebox(0,0)[lb]{\smash{{\SetFigFont{17}{20.4}{\rmdefault}{\mddefault}{\updefault}{\color[rgb]{0,0,0}$T_1$}%
}}}}
\put(4951,-2086){\makebox(0,0)[lb]{\smash{{\SetFigFont{17}{20.4}{\rmdefault}{\mddefault}{\updefault}{\color[rgb]{0,0,0}$T_2$}%
}}}}
\put(4951,-2536){\makebox(0,0)[lb]{\smash{{\SetFigFont{17}{20.4}{\rmdefault}{\mddefault}{\updefault}{\color[rgb]{0,0,0}$T_3$}%
}}}}
\put(4951,-3511){\makebox(0,0)[lb]{\smash{{\SetFigFont{17}{20.4}{\rmdefault}{\mddefault}{\updefault}{\color[rgb]{0,0,0}$T_1$}%
}}}}
\put(4951,-4786){\makebox(0,0)[lb]{\smash{{\SetFigFont{17}{20.4}{\rmdefault}{\mddefault}{\updefault}{\color[rgb]{0,0,0}$T_1$}%
}}}}
\end{picture}%